\documentclass[prl,aps,twocolumn,floatfix]{revtex4}
\usepackage{tabularx}
\usepackage{multirow}
\usepackage{array}
\usepackage{amsmath}
\usepackage{amssymb}
\usepackage{hyperref}
\usepackage{graphicx}
\usepackage{dcolumn}
\usepackage{color}
\usepackage{ulem}

\begin{document}

\title{Dicke-type phase transition in a spin-orbit coupled Bose-Einstein
condensate}
\author{Chris Hamner$^{1}$}
\author{Chunlei Qu$^{2}$}
\author{Yongping Zhang$^{1,3}$}
\author{JiaJia Chang$^{1}$}
\author{Ming Gong$^{2,4}$}
\author{Chuanwei Zhang$^{2,1}$}
\thanks{Correspondence and requests for materials should be addressed to C.Z. (email: chuanwei.zhang@utdallas.edu) or P.E. (email: engels@wsu.edu).}
\author{Peter Engels$^{1}$}
\thanks{Correspondence and requests for materials should be addressed to C.Z. (email: chuanwei.zhang@utdallas.edu) or P.E. (email: engels@wsu.edu).}

\begin{abstract}
\textbf{Spin-orbit coupled Bose-Einstein condensates (BECs) provide a
powerful tool to investigate interesting gauge-field related phenomena. We
study the ground state properties of such a system and show that it can be
mapped to the well-known Dicke model in quantum optics, which describes the
interactions between an ensemble of atoms and an optical field. A central
prediction of the Dicke model is a quantum phase transition between a
superradiant phase and a normal phase. Here we detect this transition in a
spin-orbit coupled BEC by measuring various physical quantities across the
phase transition. These quantities include the spin polarization, the
relative occupation of the nearly degenerate single particle states, the
quantity analogous to the photon field occupation, and the period of a
collective oscillation (quadrupole mode). The applicability of the Dicke
model to spin-orbit coupled BECs may lead to interesting applications in
quantum optics and quantum information science. }
\end{abstract}

\affiliation{$^{1}$Department of Physics and Astronomy, Washington State University,
Pullman, WA 99164, USA \\
$^{2}$Department of Physics, The University of Texas at Dallas, Richardson,
TX 75080, USA \\
$^{3}$Quantum System Unit, Okinawa Institute of Science and Technology,
Okinawa 904-0495, Japan \\
$^{4}$ Department of Physics and Center of Coherence, The Chinese University
of Hong Kong, Shatin, N.T., Hong Kong, China}
\maketitle

Ultracold atomic gases afford unique opportunities to simulate
quantum-optical and condensed matter phenomena, many of which are difficult
to observe in their original contexts ~\cite{Bloch1,Bloch2}. Over the past
decade, much theoretical and experimental progress in implementing quantum
simulations with atomic gases has been achieved, exploiting the flexibility
and tunability of these systems. The recent generation of spin-orbit (SO)
coupling in Bose-Einstein Condensates (BECs)~\cite{Ian,Pan,Peter,Chen} and
Fermi gases~\cite{Zhang,Martin,Ian4} has brought the simulation of a large
class of gauge field related physics into reach, such as the spin Hall
effect~\cite{Ian2,Ian3, Bloch3,Ketterle}. With such achievements, SO coupled
ultracold atomic gases have emerged as excellent platforms to simulate
topological insulators, topological superconductors/superfluids etc., which
have important applications for the design of next-generation spin based
atomtronic devices and for topological quantum computation~\cite{Kane,Qi}.

Recently, the ground state properties of a BEC with one dimensional (1D) or
two dimensional (2D) SO coupling have been analyzed theoretically. These
investigations have predicted a plane wave or stripe phase for different
parameter regimes~\cite{Zhai,Wu,Ho2,Sinha,Yongping2,Hu,Stringari}, agreeing
with the experimental observations \cite{Ian}.  In the plane wave phase of
such a SO coupled BEC, the atomic spins collectively interact with the
motional degrees of freedom in the external trapping field, providing a
possible analogy to the well known quantum Dicke model. The Dicke model
\cite{Dicke}, proposed nearly sixty years ago, describes the interaction
between an ensemble of two-level atoms and an optical field~\cite{Esslinger}%
. For atom-photon interaction strengths greater than a threshold value, the
ensemble of atoms favors to interact with the optical field collectively as
a large spin and the system shows an interesting superradiant phase with a
macroscopic occupation of photons and non-vanishing spin polarization \cite%
{Hepp,Wang,Gross}. Even though this model has been solved and is well
understood theoretically, the experimental observation was achieved only
recently by coupling a BEC to an optical cavity~\cite{Baumann}.

\begin{figure}[tbp]
\centering
\includegraphics[width=3.2in]{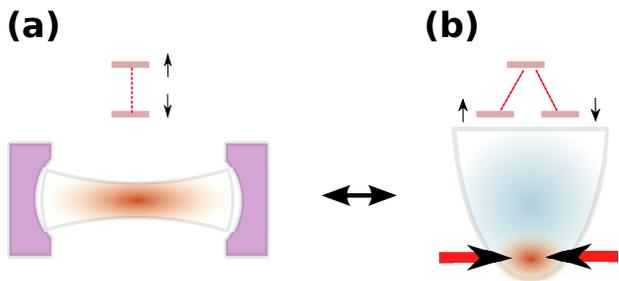}
\caption{\textbf{Analogy between standard Dicke model and SO coupled BEC.}
(a) Standard Dicke model describing the interaction of an ensemble
of two-level atoms in an optical cavity. The optical model in the cavity
couples two atomic spin states. (b) SO coupled BEC in an
external trap. Two spins states are coupled by two counter-propagating Raman
lasers. }
\label{Fig1}
\end{figure}

In this work we experimentally investigate the ground state properties of
the plane wave phase of a SO coupled BEC and show that an insightful
analogy to the quantum-optical Dicke model can be constructed. The SO
coupling in a BEC is realized with a Raman dressing scheme. The system
exhibits coupling between momentum states and the collective atomic spin
which is analogous to the coupling between the photon field and the atomic
spin in the Dicke model. This analogy is depicted in Fig.~\ref{Fig1}. By
changing the Raman coupling strength, the system can be driven across a
quantum phase transition from a spin-polarized phase, marked by a non-zero
quasi-momentum, to a spin balanced phase with zero quasi-momentum, akin to
the transition from superradiant to normal phases in Dicke model.
Measurements of various physical quantities in these two phases are
presented.

{\LARGE \textbf{Results}}

\textbf{Theoretical description of the SO coupled BEC:} The Raman dressing
scheme is based on coupling two atomic hyperfine states in such a way that a
momentum transfer of $2\hbar k_{\text{R}}$ in the $x$ direction is
accompanied with the change of the hyperfine states, where $\hbar k_{\text{R}%
}$ is the photon recoil momentum. The dynamics in the $y$ and $z$ directions
are decoupled, which allows us to consider a 1D system in our following
discussions (see Methods). The two coupled hyperfine states are regarded as
the two orientations of a pseudo-spin $1/2$ system. The Raman dressed BEC is
governed by the 1D Gross-Pitaevskii (G-P) equation with the Hamiltonian $H_{%
\text{SO}}=H_{\text{s}}+H_{\text{I}}$. Here $H_{\text{s}}$ is the single
particle Hamiltonian and in the basis of the uncoupled states can be written
as
\begin{equation}
H_{\text{s}}=%
\begin{pmatrix}
\frac{\hbar ^{2}}{2m}(k_{x}+k_{\text{R}})^{2}+{\frac{\delta }{2}} & {\frac{%
\Omega }{2}} \\
{\frac{\Omega }{2}} & \frac{\hbar ^{2}}{2m}(k_{x}-k_{\text{R}})^{2}-{\frac{%
\delta }{2}}%
\end{pmatrix}%
+V_{\text{t}}\ {.}  \label{eq-single}
\end{equation}%
$\Omega $ is the Raman coupling strength, and $\delta $ is the detuning of
the Raman drive from the level splitting. The recoil energy is defined as $%
E_{\text{R}}=\hbar ^{2}k_{\text{R}}^{2}/{2m}$. $k_{x}$ is the
quasi-momentum, and $V_{\text{t}}=m\omega _{x}^{2}x^{2}/2$ is the external
harmonic trap. The many-body interactions between atoms are described by
\begin{equation}
H_{\text{I}}=\text{diag}\left( \sum\nolimits_{\sigma =\uparrow ,\downarrow
}g_{\uparrow \sigma }|\psi _{\sigma }|^{2},\sum\nolimits_{\sigma =\uparrow
,\downarrow }g_{\downarrow \sigma }|\psi _{\sigma }|^{2}\right) {,}
\label{eq-nonlinear}
\end{equation}%
where $g_{\alpha \beta }$ are the effective 1D interaction parameters (see
Methods) \cite{Ho,Bloch4}. The presence of the interatomic interactions is
crucial to observe the Dicke phase transition. For $^{87}$Rb atoms, the
differences between the spin dependent nonlinear coefficients are very small
and contribute only small modifications to the collective behavior (see
Methods).

The band structure of the non-interacting system with $\Omega<4E_\text{R}$
and $\delta=0$ has two degenerate local minima at quasi-momenta $\pm q$,
where $q= k_{\text{R}}\sqrt{1-(\Omega/{4E_\text{R}})^2}$. The spin
polarization of these two states is finite and opposite to each other. An
ensemble of non-interacting atoms occupies both states equally and thus has
zero average spin polarization and quasi-momentum. When the nonlinear
interactions are taken into account, a superposition state with components
located at both degenerate minima generally has an increased energy and is
thus not the many-body ground state \cite{StamperKurn}. The ground state of
the BEC is obtained when the atoms occupy one of the degenerate single
particle ground states (L or R). This is depicted in the inset of Fig.~\ref%
{Fig2}a.

The mean-field energy associated with a spin-flip in an interacting,
harmonically trapped BEC is determined by the coupling between the atomic
spin and the many-body ground state harmonic mode. This situation is similar
to that of many two-level atoms interacting with a single photon field in an
optical cavity~\cite{Dicke}. As shown in the following, the interaction
induced spontaneous symmetry breaking of the ground state allows the mapping
of the SO coupled BEC to the well-known Dicke model, leading to the
prediction of an intriguing quantum phase transition when the Raman coupling
strength is varied~\cite{Yongping}.

\begin{figure}[tbp]
\centering
\includegraphics[width=3.2in]{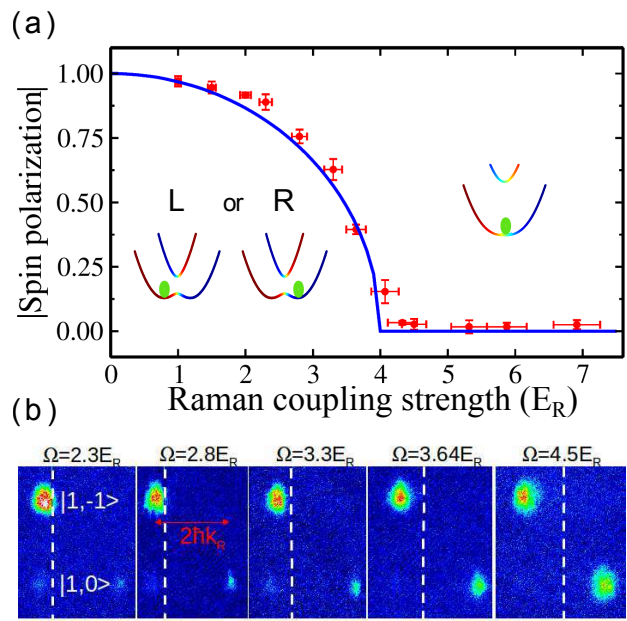}
\caption{\textbf{Spin polarization and corresponding quasi-momentum from
time-of-flight image.} (a) Absolute value of the spin polarization as a
function of Raman coupling strength $\Omega $ for $\protect\delta =0$. The
solid blue line gives the spin polarization predicted by the Dicke model.
The symbols are the experimentally measured data. The vertical error bars
are the standard deviation for 4 to 5 realizations, while the horizontal
error bars reflect the systematic uncertainties in the determination of $%
\Omega$. The insets are examples of the dispersion relation in the regime of
$\Omega $ below or above $4E_{\text{R}}$, and the color of the bands
represents the spin composition. (b)Example of experimental time-of-flight
images of BECs loaded to $+q$. The pseudo-spin states are horizontally
separated by $2\hbar {k_{\text{R}}}$ due to the Raman momentum transfer and
the dashed vertical line indicates zero momentum. The separation of the spin
states in the vertical direction is achieved by a Stern-Gerlach field which
is briefly applied during time-of-flight.}
\label{Fig2}
\end{figure}

\textbf{Mapping to the Dicke model: }The realization of the Dicke transition
requires two conditions: a single mode that interacts with all atoms and the
thermodynamic (large atom number) limit. While the latter is naturally
fulfilled in our spin-orbit coupled BECs with $N>10^{4}$ atoms, the first
condition is driven by the many-body interaction between atoms. Note that
the mean-field interaction energy contains two types of terms (see Methods):
a term proportional to the overall interaction strength ($\propto
g_{\uparrow \uparrow }+g_{\downarrow \downarrow }+2g_{\uparrow \downarrow }$%
) that prefers a uniform density (the plane wave phase), and a term
proportional to the difference of the interaction strengths in different
spin states $\left( \propto g_{\uparrow \uparrow }-g_{\uparrow \downarrow
}\right) $ that prefers spatial modulation of the density. In the
experiment, the first term dominates for a reasonably large $\Omega $, and
the plane wave phase is preferred (i.e., possesses lowest energy) \cite%
{Stringari}. In this phase, all atoms collectively interact with a single
plane wave mode, similar to atoms confined in a cavity interacting only with
the cavity mode.

To see the connection between the SO coupled BEC and the Dicke model, the
interacting many-body ground state is first expressed in terms of the
harmonic trap mode (see Methods). Setting $p_{x}=\hbar k_{x}=i\sqrt{m\omega
_{x}\hbar/2 }(a^{\dagger }-a)$, the $N$-particle Hamiltonian can be written
as
\begin{eqnarray}
H_{\text{Dicke}} &=&N\hbar \omega _{x}a^{\dagger }{a}+ik_{R}\sqrt{\frac{%
2\hbar \omega _{x}}{m}}(a^{\dagger }-a)J_{z}+\frac{\Omega }{\hbar }J_{x}
\notag \\
&+&(\frac{4G_{3}}{N\hbar }+\frac{\delta }{\hbar })J_{z}+\frac{4G_{3}}{%
N^{2}\hbar ^{2}}J_{z}^{2}+\text{const}.  \label{eq-Dicke}
\end{eqnarray}%
where the uniform approximation has been adopted to treat the nonlinear
interaction term, $G_{3}=n(g_{\uparrow \uparrow }-g_{\downarrow \downarrow
})/4 $, $n$ is the local density. $J_{x,z}$ are the collective spin
operators defined as $J_{x}={\hbar/2}\sum \sigma _{x}^{i},J_{z}=\hbar/2 \sum
\sigma _{z}^{i}$. $a^{\dagger }a$ is the occupation number of the harmonic
trap mode. The differences between the interaction energies contribute an
effective detuning term and a nonlinear term in the large spin operator, $%
J_{z}^{2}$. However, these terms are small for the experimental states
chosen and thus are ignored in the following analysis. For $\delta =0$, the
Hamiltonian of the first line in Eq.~\ref{eq-Dicke} is equivalent to the
Dicke model~\cite{Dicke}. A quantum phase transition between the normal
phase and a superradiant phase can be driven by changing the Raman coupling
strength $\Omega $.

The critical point for the phase transition can be derived using the
standard mean-field approximation \cite{Emary,Emary2,Bakemeier} yielding $%
\Omega _{\text{c}}=4E_{\text{R}}$ (note that the $J_{z}^{2}$ term yields a
small correction $-4G_{3}/N$ to $\Omega _{c}$, which is neglected here).
When $\Omega <\Omega _{\text{c}}$, the Dicke model predicts that the
dependence of the order parameter on the Raman coupling strength is $\langle
\sigma_{z}\rangle = \langle J_{z}\rangle /j\hbar =\pm \sqrt{1-\Omega ^{2}/{%
16E_{\text{R}}^{2}}}$. For $\Omega >\Omega _{\text{c}}$, one obtains $%
\langle J_{z}\rangle /j\hbar =0$ and $\langle \sigma _{x}\rangle =\langle
J_{x}\rangle /j\hbar =1 $ ($j=N/2$). This scaling is confirmed in our
numerical simulation of the G-P equation. The ground state of the BEC is
obtained through an imaginary time evolution. The spin polarization can be
calculated as $|\langle \sigma _{z}\rangle |=\left\vert \int dx\left(
\left\vert \psi _{\uparrow }\right\vert ^{2}-\left\vert \psi _{\downarrow
}\right\vert ^{2}\right) \right\vert $. The absolute value of $\langle
\sigma _{z}\rangle $ is taken since $\pm q$ are spontaneously chosen. The
scaling of $|\langle J_{z}\rangle |/(j\hbar )$ is shown in Fig.~\ref{Fig2}
and clearly is consistent with the experimental data described in the
following.

\textbf{Experimental procedure:} To experimentally probe this system, we
adiabatically Raman dress a BEC of $^{87}$Rb atoms in the $\left\vert
\downarrow \right\rangle \equiv |F=1,m_{F}=-1\rangle $ and $\left\vert
\uparrow \right\rangle \equiv |F=1,m_{F}=0\rangle $ hyperfine states (see
Methods). A magnetic bias field is applied that generates a sufficiently
large quadratic Zeeman splitting such that the $|F=1,m_{F}=1\rangle $ state
can be neglected. The system can thus be treated as an effective two state
system. To analyze the bare state composition, all lasers are switched off
and the atoms are imaged after time-of-flight in a Stern-Gerlach field. This
separates the bare states along the vertical axis of the images. The
absolute value of the spin polarization, given by $|(N_{\uparrow
}-N_{\downarrow })|/(N_{\uparrow }+N_{\downarrow })$, and the quasi-momentum
are directly measured for condensed atoms at $\pm q$. The experimentally
measured absolute value of the spin polarization for various Raman coupling
strengths $\Omega $ spanning the quantum phase transition is shown in Fig.~%
\ref{Fig2}a.

\begin{figure}[tbp]
\centering
\includegraphics[width=3.2in]{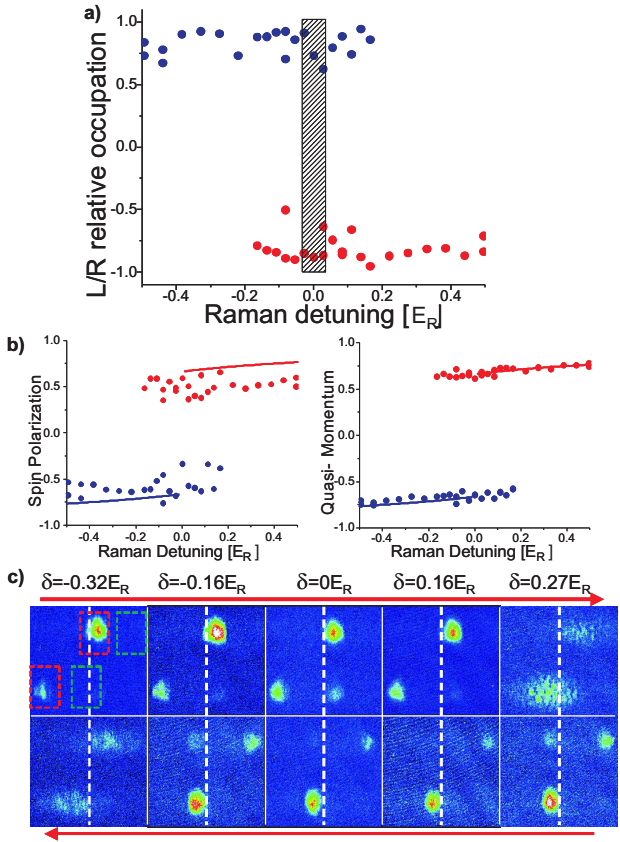}
\caption{\textbf{Effect of a small detuning on the superradiant
phase.}  (a) Experimentally measured relative occupation of the two
dispersion minima in the superradiant phase. The system is loaded
adiabatically for $\Omega =3~E_{\text{R}}$ at a detuning of $-5.4~E_{\text{R}%
}$ (blue points) or $+5.4~E_{\text{R}}$ (red points). Then the detuning is
linearly swept to the final value in $50~\text{ms}$. The hashed region
indicates the experimental uncertainty to which $\protect\delta =0$ can be
determined. (b) Experimental measurements of the spin polarization and
quasi-momentum, overlaid with the single particle expectations
(lines). (c) Absorption images taken after projecting the BEC onto the bare
states, applying Stern-Gerlach separation, and $11.5~\text{ms}$ time of
flight. The given values of $\protect\delta$ are the endpoints of the
detuning ramps described in the main text. The relative occupation in (a) is
calculated from the atom numbers contained in the dashed boxes in (c). The
dashed vertical line indicates zero kinetic momentum, while the horizontal
red arrows indicate the direction in which $\protect\delta $ is swept.}
\label{Fig3}
\end{figure}

An experimental investigation of the collective choice of the atoms to
occupy a single minimum in the dispersion relation is presented in Fig.~\ref%
{Fig3}. For this data, the BEC is adiabatically prepared in a Raman dressed
state with $\Omega =3~E_{\text{R}}$ and a detuning $\delta$ of either $\pm
5.4~E_{\text{R}}$. The detuning is then linearly swept to a final value in
50~ms. The left/right relative occupation of the BEC, defined by $%
(N_{+q}-N_{-q})/(N_{+q}+N_{-q})$, provides a measure for the relative
occupation of the dressed states at the quasi-momentum of the two dispersion
minima. The data presented in Fig.~\ref{Fig3} exhibits a hysteretic effect
dependent on the sign of the initial detuning $\pm \delta $. The width of
the hysteresis depends upon the chosen sweep rate of the detuning (see
Methods). While these sweeps are slow enough for the BEC to follow a chosen
minimum of the band structure, the transition between the two possible
minima is non-adiabatic. Sweeps causing a reversal of the energetic order of
the two minima eventually generate heating of the BEC, shown in Fig.~\ref%
{Fig3}c. Only values of $\delta $ where the BEC is not heated are plotted
in Fig~\ref{Fig3}a,b). Recently, a similar energy dispersion exhibiting a
double well structure was also observed for a BEC in a shaken optical
lattice, where the analogous spontaneous occupation of a single band mimimum
was also found ~\cite{Chin}. However, such a system lacks the linear
coupling between momentum and spin presented in our system (Eq. \ref%
{eq-single}), and thus cannot be mapped to the Dicke model.

\begin{figure}[tbp]
\centering
\includegraphics[width=3.0in]{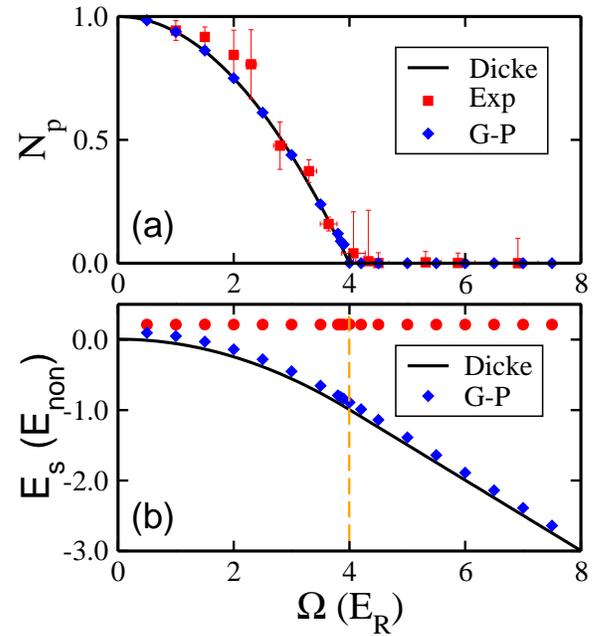}
\caption{\textbf{Photon field excitation and ground state energy.} (a)
Scaled photon field excitation of the effective Dicke model. The red symbols
are the experimental results from the quasi-momentum measurements. The
vertical error bars are the standard deviation for 4 to 5 realizations,
while the horizontal error bars reflect the systematic uncertainties in the
determination of $\Omega $. The blue symbols are the simulation results
based on the G-P equation, and the solid line is the prediction of the Dicke
model. (b) Numerical results for the single particle ground state energy $%
E_{s}$ (blue diamonds) and the nonlinear interaction energy $E_{non}$ (red circles) as
a function of Raman coupling strength. The black solid line shows
the single particle energy from the prediction of the Dicke model. The
dashed vertical line indicates the location of the quantum phase transition.}
\label{Fig4}
\end{figure}

From the effective Dicke model, we also obtain the quantity analogous to the
photon field excitation, as well as the ground state energy of the system.
The photon number $N_{\text{p}}$ is related to the quasi-momentum $q$ of the
SO coupled BEC: $N_{\text{p}}=q^{2}$ after a constant coefficient is scaled
to unity (see Methods). As seen from the experimental measurements and
numerical simulations in Fig.~\ref{Fig4}a, the superradiant phase has a
macroscopic photon excitation, while the photon field vanishes in the normal
phase. The ground state energy $E_{\text{s}}$ of a single particle, as
obtained from G-P simulations, is plotted in Fig.~\ref{Fig4}b and is in
agreement with the predictions of the Dicke model. The second order
derivative of $E_{\text{s}}$ is discontinuous at the critical point $\Omega
=4E_{\text{R}}$, reinforcing the nature of the quantum phase transition. The
red circles in Fig.~\ref{Fig4}b show the density dependent mean field
interaction energy in the numerical simulations, contributing approximately $%
0.2E_\text{R}$ to the total energy.

\begin{figure}[tbp]
\includegraphics[width=3.2in]{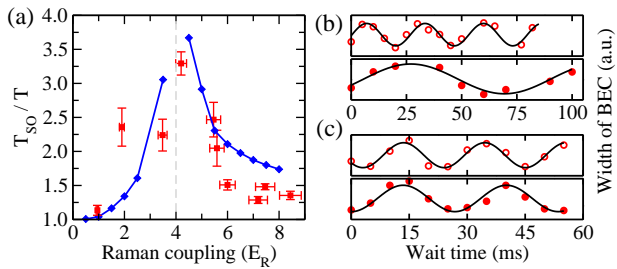}
\caption{\textbf{Quadrupole excitations across the Dicke
phase transition.} (a) Scaled quadrupole oscillation period for $\protect%
\delta \approx 0$. The blue points and the solid blue line are the results
of numerical simulations. The dashed vertical line indicates the location of
the quantum phase transition. The red symbols with error bars are the
experimentally measured data. The vertical error bars are the uncertainty of
the oscillation frequency from the sinusoidal fits, while the horizontal
error bars reflect the systematic uncertainties in the determination of $%
\Omega $. (b,c) Experimentally observed temporal oscillation of the
condensate width for (b) $\Omega =4.2~E_{\text{R}}$ and (c) $\Omega =~7.2E_{%
\text{R}}$. The upper and lower panel of each figure represents the
off-resonant case ($\protect\delta >>\Omega $) and on-resonant case ($%
\protect\delta \approx 0$), respectively. The solid lines are fits to the
experimental data.}
\label{Fig5}
\end{figure}

\textbf{Collective excitations:}  It is well known that various physical quantities may
change dramatically across a quantum critical point~\cite{Sachdev}. As a
particular example, we investigate the quadrupole collective excitation of a
SO coupled BEC. In BECs without SO coupling, the quadrupole excitation
frequency only depends on the trapping geometry and on the ratio of the
kinetic energy to the trapping energy \cite{Stringari2}. Measurements and
G-P simulations of the quadrupole mode frequencies for SO coupled BECs are
shown in Fig.~\ref{Fig5} as a function of the Raman coupling strength. For
this data, the quadrupole oscillation period is scaled by the period
measured for an off resonant case ($\delta >> \Omega$). This removes a
dependence on the trapping geometry which changes with the Raman coupling
strength in the experiment. A peak in the oscillation period around the
phase transition is observed. In the G-P simulations, a quadrupole
oscillation amplitude of approximately $0.2\hbar k_{\text{R}}$ was used,
whereas the oscillation amplitude in the experimental results varied from $%
0.15~\hbar k_{\text{R}}$ to $0.55~\hbar k_{\text{R}}$. In SO coupled
systems, the hydrodynamic mode frequencies depend strongly on the
oscillation amplitude \cite{Pan}. This may contribute to the variation of
the experimental data. The numerical results reveal that the oscillations of
the BEC are undamped in both spin balanced and polarized phases, but show
strong damping in the transition region. The behavior of the quadrupole mode
provides an additional signature of the quantum phase transition. Similar
experimental behavior has been observed for the collective dipole motion \cite{Pan}.

\begin{figure}[tbp]
\centering
\includegraphics[width=3.0in]{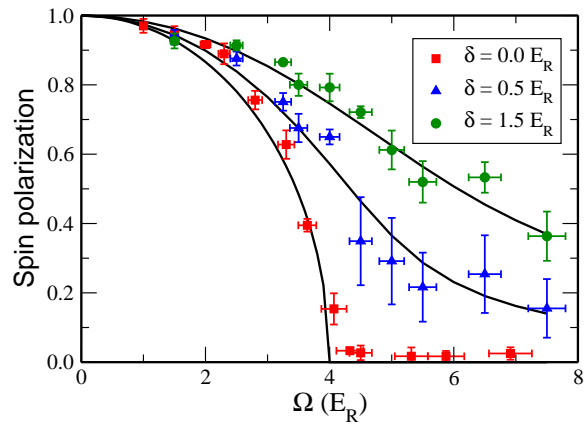}
\caption{\textbf{Realization of a generalized Dicke model.} Absolute value of the spin
polarization as a function of Raman coupling strength $\Omega $ for detunings $\protect\delta =0~E_{\text{R}}$, $0.5~E_{\text{R}}$ and $1.5~E_{\text{R}}$. For a finite detuning there is only one non-degenerate solution of the
generalized Dicke model, and no quantum phase transition exists. The data
points are experimental measurements and the solid lines are the
corresponding theoretical predictions from the generalized Dicke model. The
vertical error bars are the standard deviation for 4 to 5 realizations,
while the horizontal error bars reflect the systematic uncertainties in the
determination of $\Omega$.}
\label{Fig6}
\end{figure}

\textbf{Finite detuning: } While the analysis above has focused on the case
of $\delta =0$, a finite detuning leads to the realization of the
generalized Dicke Hamiltonian \cite{Emary3}. For finite detuning, the sharp
quantum phase transition vanishes and the BEC is always in a superradiant
(spin polarized) phase, as shown in Fig.~\ref{Fig6}. The inclusion of this
parameter demonstrates the completeness in the mapping of the SO coupled BEC
to a Dicke type system.

{\LARGE \textbf{Discussion}}

We have demonstrated that varying the Raman coupling strength
can drive a harmonically trapped SO coupled BEC across a quantum phase
transition from a spin balanced to a spin polarized ground state. This is a
realization of the long-sought phase transition from a normal to a
superradiant phase in the quantum Dicke model. The corresponding spin
polarization and photon field occupation are calculated for the Dicke model
and agree with the experimental measurements and the numerical simulations.
The ground state energy is found to be consistent with our numerical
simulations as well. Using SO coupled BECs to study Dicke model physics may
have important applications in quantum information and quantum optics
including spin squeezing, quantum entanglement, etc. \newline

{\LARGE \textbf{Methods}}

\textbf{Experimental setup and parameters:} For the experiments presented in
this manuscript, the BECs are held in a crossed optical dipole trap with
frequencies $\{\omega_x, \omega_y,\omega_z\}$ within the range of $%
2\pi\times\{12-34~\text{Hz}, 134~\text{Hz}, 178~\text{Hz}\}$ respectively.
Depending on the Raman coupling strength and crossed dipole trap potential,
BECs of $2\times 10^4$ to $10\times 10^4$ atoms are loaded into the dressed
state. Small atom numbers are favorable in this trapping geometry in order
to reduce collisions between atoms differing by $2\hbar k_\text{R}$ during
time-of-flight. The Raman beams, intersecting under an angle of $\pi/2$, are
individually oriented at an angle of $\pi/4$ from the long axis of the BEC.
They are operated between the D1 and D2 lines of $^{87} $Rb, in the range of
$782.5-790~\text{nm}$. A $10~\text{G}$ magnetic bias field is applied along
the long axis of the BEC and produces a quadratic Zeeman shift of
approximately $7.4~E_\text{R}$. The BEC is imaged after performing
Stern-Gerlach separation during a 11.5~ms time-of-flight expansion. The
images directly reveal the spin and momentum distributions.

For each value of $\Omega$, $\delta=0$ is experimentally identified by
finding the Raman laser detuning for which the spin polarization is
minimized (normal phase) or for which the BEC switches between the two
degenerate states (super-radiant phase). This method compensates for the
presence of the third atomic hyperfine state ($|1, +1\rangle$) in the F=1
manifold. The magnetic bias field was actively stabilized using a technique
similar to the one described in Ref.~\cite{truscott}.

\textbf{Experimental preparation methods:} To experimentally measure the
absolute value of the spin polarization $|\langle \sigma _{z}\rangle |$ and
the photon field occupation (shown in Fig.~\ref{Fig2} and Fig.~\ref{Fig4}),
we start with a BEC in the bare state $\left\vert \uparrow \right\rangle $ ($%
\left\vert \downarrow \right\rangle $) and adiabatically turn on the Raman
dressing at $\delta =5.4E_{\text{R}}$ ($-5.4E_{\text{R}}$). The initial
state and the dressed state have similar spin compositions. This is followed
by a linear sweep of the detuning to a final value in $50-100~\text{ms}$. An
example of results obtained with this loading method is shown in Fig.~\ref%
{Fig1Supp} for the normal phase of the Dicke model with $\Omega =6E_{\text{R}%
}$. This loading method produces BECs with only minimal excitations when the
final detuning value is chosen near $\delta=0$. As seen in Fig.~\ref{Fig3}
this method can cause a hysteresis like feature at small $|\delta |$.

\begin{figure}[bp]
\centering
\includegraphics[width=3.2in]{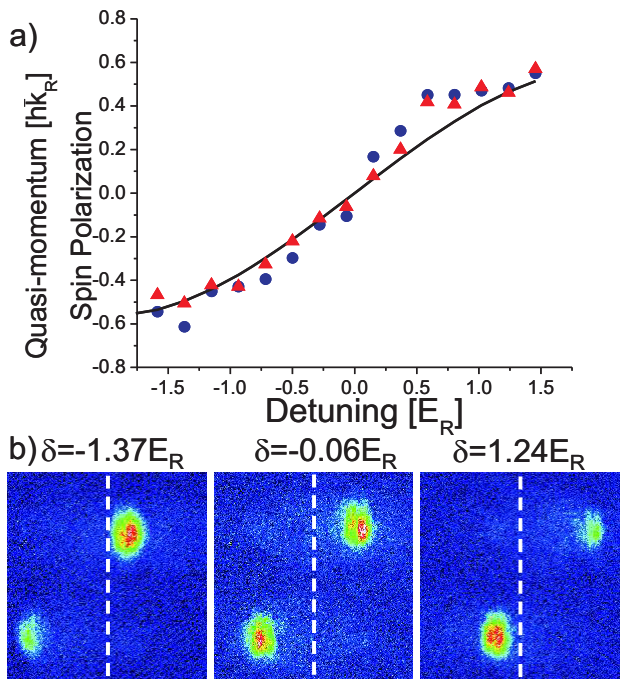}
\caption{\textbf{Effect of a small detuning on the normal phase.}  (a) Spin
polarization (triangles) and quasi-momentum (circles) measured for the
normal phase, where there is no degeneracy in the ground state of the single
particle dispersion. This measurement was performed at $\Omega =6E_{\text{R}}
$ for various $\protect\delta $. The data is overlaid with the single
particle expectation for the ground state of the system (solid line). (b)
Experimental images for three different realizations in (a). The dashed
vertical line indicates zero momentum.}
\label{Fig1Supp}
\end{figure}

\begin{figure}[tbp]
\centering
\includegraphics[width=3.2in]{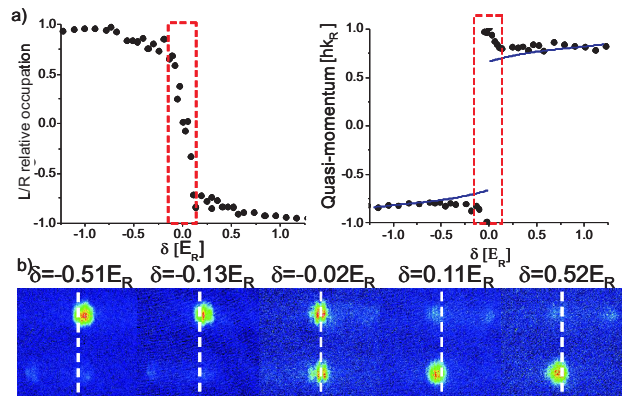}
\caption{ \textbf{Effect of a small detuning on the superradiant phase using an alternative loading method.} 
(a) Experimental determination of the L/R relative occupation of the BEC in
the superradiant phase for $\Omega =3E_{\text{R}}$. The Raman coupling is
ramped on over 60 ms at a fixed laser detuning (which differs from the
condition of fixed $\protect\delta $) followed by a 150~ms wait time. For
all but small detunings the system is loaded to a single point in the
dispersion relation. (b) The accompanying quasi-momentum deviates from the
expected single particle value at small detunings where the BEC acquires excitations. This region is highlighted by the dashed
red boxes. (c) Experimental images.}
\label{Fig2Supp}
\end{figure}

An alternative loading method, as opposed to ramps of $\delta$, is
investigated by fixing the laser detuning while $\Omega$ is increased. The
resulting L/R relative occupation measurement is shown in Fig.~\ref{Fig2Supp}%
. Here the BEC is loaded to the desired parameters in approximately $60~%
\text{ms}$ by only increasing the Raman beam intensities. This is followed
by a $150~\text{ms}$ wait time, allowing for excitations to be dissipated.
While this wait time is long, there may still be some residual small
amplitude collective excitations, such as dipole oscillations. When this
technique is used in Fig.~\ref{Fig2Supp}, the BEC occupies only one of the
two minima for all but small detunings. In the range where the BEC fails to
load to a single point in the dispersion relation, the quasi-momentum (shown
in Fig.~\ref{Fig2Supp}b, strongly deviates from the single
particle expectation and the BEC acquires excitations. While this loading
method does not result in the occupation of the ground state at $\delta=0$,
the BEC does choose a single minimum for detunings much smaller than the
chemical potential $\mu$, which is on the order of $0.55 E_\text{R}$ for the
data shown. Similar results are obtained by adding sufficient wait time
following the linear sweeps of $\delta$ performed for Fig.~\ref{Fig3}. A
related effect has been observed in \cite{Pan2}.

The quadrupole frequency measurements begin with a BEC in the $%
|F,m_{F}\rangle =|1,-1\rangle $ state in the crossed dipole trap. The two
Raman beam intensities are linearly increased, one after the other, over $%
30-150~\text{ms}$. This loading method, performed with the Raman beams at $%
782.5~\text{nm}$, generates large quadrupole and small dipole excitations.
To simplify the analysis only the first few oscillations of the quadrupole
mode are fit to an undamped oscillation. In the numerical simulations the
quadrupole mode is excited by a sudden jump of the trap frequency along the
SO coupling direction. Evolution time is added to observe the temporal
oscillation of the condensate width which reveals the quadrupole mode
frequency.

\textbf{Dimension reduction from 3D to 1D:} The dynamics of SO coupled BECs
and their mapping to the Dicke model are studied based on the 1D G-P
equation, although the experimental system is three dimensional with strong
confinements along two directions. The SO coupling is along the
elongated direction of the trap, therefore the transverse degrees of freedom
do not couple with the internal spin states. Assuming the harmonic ground
states along the transverse directions, the effective 1D nonlinear
interaction coefficients $g_{\text{1D}}$ can be approximately obtained from
3D through $g_{\text{1D}}=g_{\text{3D}}/l_{y}l_{z}$, where $l_{y,z}=\sqrt{%
\hbar/m\omega_{y,z}}$ are the harmonic characteristic lengths along the
transverse directions, and $g_{\text{3D}}=4N\pi\hbar^2a_\text{s}/m$ is the
3D nonlinear interaction coefficient. For the study of the quadrupole mode,
we simulate the 2D G-P equation in the $xy$ plane to compare with the
experimental observations, where the dimension reduction gives $g_{\text{2D}%
}=g_{\text{3D}}/l_{z}$.

\textbf{Mapping to the Dicke-type Hamiltonian:} In the mean field
approximation, the single-particle Hamiltonian is given by
\begin{equation}
H_{\text{s}}=\frac{p_{x}^{2}}{2m}+\frac{1}{2}m\omega _{x}^{2}x^{2}+\frac{%
\hbar k_{\text{R}}}{m}p_{x}\sigma _{z}+\frac{\Omega }{2}\sigma _{x}+\frac{%
\delta }{2}\sigma _{z}+E_{\text{R}}.
\end{equation}%
As discussed in the main text, the large nonlinear interactions enable the
atoms to collectively occupy the same many-body ground state, forming a
single mode. For an ensemble of $N$ interacting bosonic atoms, we define the
collective spin operators $J_{x}=\hbar/2 \sum_{i}\sigma _{x}^{i}$ and $%
J_{z}=\hbar/2 \sum_{i}\sigma _{z}^{i}$. Substituting these collective
operators into the $N$-particle Hamiltonian and using the harmonic mode
operator, $p_{x}=i\sqrt{m\omega_{x}{\hbar }/2}(a^{\dagger }-a)$, we obtain
\begin{eqnarray}
H_{\text{s}}&=&N\hbar \omega _{x}(a^{\dagger }{a}+\frac{1}{2})+ik_{\text{R}}%
\sqrt{\frac{2\hbar \omega _{x}}{m}}(a^{\dagger }-a)J_{z}+\frac{\Omega }{%
\hbar }J_{x}  \notag \\
&+& \frac{\delta }{\hbar }J_{z}+NE_{\text{R}}{,}
\end{eqnarray}%
which corresponds to the generalized Dicke model.

The interaction term in the mean field approximation is
\begin{equation*}
H_{\text{I}}=%
\begin{pmatrix}
g_{\uparrow \uparrow }|\psi _{\uparrow }|^{2}+g_{\uparrow \downarrow }|\psi
_{\downarrow }|^{2} & 0 \\
0 & g_{\downarrow \uparrow }|\psi _{\uparrow }|^{2}+g_{\downarrow \downarrow
}|\psi _{\downarrow }|^{2}%
\end{pmatrix}%
,
\end{equation*}%
which can also be formally expressed using collective operators. We define
the variables $G_{1}=n(g_{\uparrow \uparrow }+g_{\downarrow \downarrow
}+2g_{\uparrow \downarrow })/8$, $G_{2}=n(g_{\uparrow \uparrow
}+g_{\downarrow \downarrow }-2g_{\uparrow \downarrow })/8$ and $%
G_{3}=n(g_{\uparrow \uparrow }-g_{\downarrow \downarrow })/4$~\cite%
{Stringari}, where $n$ is the average density of the BEC. For the parameters
used in our experiment, we have $g_{\uparrow \downarrow }=g_{\downarrow
\downarrow }\approx g_{\uparrow \uparrow }$, thus $G_{3}=2G_{2}\approx 0$
and $G_{1}\approx n{g_{\uparrow \uparrow }}/2$.

Because of the normalization condition of the BEC wavefunction, it is easy
to see that $|\psi _{\uparrow }|^{2}+|\psi _{\downarrow }|^{2}=n/N$, and $%
|\psi _{\uparrow }|^{2}-|\psi _{\downarrow }|^{2}=2nJ_z/N^2\hbar $ ($%
J_{z}=N\hbar /2$ when all atoms are in the spin up state). With these
notations, the Hamiltonian can be mapped to
\begin{eqnarray}
H_{\text{Dicke}} &=&N\hbar \omega _{x}a^{\dagger }{a}+ik_{\text{R}}\sqrt{%
\frac{2\hbar \omega _{x}}{m}}(a^{\dagger }-a)J_{z}+\frac{\Omega }{\hbar }%
J_{x}+2G_{1}  \notag \\
&+&(\frac{4G_{3}}{N\hbar }+\frac{\delta }{\hbar })J_{z}+\frac{4G_{3}}{%
N^{2}\hbar ^{2}}J_{z}^{2}+\frac{N\hbar \omega _{x}}{2}+NE_{\text{R}}.
\end{eqnarray}
Note that $g_{\alpha\beta}=4N\pi\hbar^2 a_{\alpha\beta}/ml_{y}l_{z}$ is
proportional to $N$, thus all the terms in the above Hamiltonian scale as $N$%
.

\textbf{Dicke phase transition} The critical point for the phase transition
of the Dicke model can be derived using the mean field coherent state \cite%
{Bakemeier}, where the mean-field ansatz of the ground state wavefunction is
given by
\begin{equation}
|\psi \rangle =|\theta \rangle \otimes |\alpha \rangle .  \notag
\end{equation}%
Here the spatial coherent state $|\alpha \rangle $ is defined by $a|\alpha
\rangle =\alpha |\alpha \rangle $, and the spin coherent state $|\theta
\rangle$ is defined as
\begin{equation}
|\theta \rangle =e^{i\theta {J_{y}/}\hbar }|j,-j\rangle  \notag
\end{equation}%
with $j=N/2$ for spin $1/2$ atoms and $\theta \in {[0,2\pi ]}$.

In the absence of the detuning term $(4G_3/N\hbar+\delta/\hbar)J_z$, the
ground state energy of the Dicke Hamiltonian is%
\begin{eqnarray}
E(\theta ,\alpha ) &=&\left\langle \psi \right\vert H_{\text{Dicke}%
}\left\vert \psi \right\rangle  \notag \\
&=&N\hbar \omega _{x}(u^{2}+v^{2})-N{\hbar }k_{\text{R}}\sqrt{\frac{2\hbar
\omega _{x}}{m}}v\cos \theta +2G_{1}  \notag \\
&+&\frac{\Omega }{\hbar }(\frac{N}{2}\hbar \sin \theta )+G_{3}\cos
^{2}\theta +\frac{N\hbar \omega _{x}}{2}+NE_{\text{R}}
\end{eqnarray}%
where $u$ and $v$ are the real and imaginary parts of $\alpha $, i.e., $%
\alpha =u+iv$. Minimizing $E(\theta ,\alpha )$ with respect to $u$ and $v$
leads to
\begin{eqnarray}
u &=&0  \label{Minimizing1} \\
v &=&k_{\text{R}}\sqrt{\frac{\hbar }{2m\omega _{x}}}\cos \theta .
\label{Minimizing2}
\end{eqnarray}%
Thus the ground state energy becomes
\begin{eqnarray}
E(\theta ,\alpha )&=&-NE_{\text{R}}\left[(1-\frac{G_{3}}{NE_{\text{R}}})\cos
^{2}\theta -\frac{\Omega }{2E_{\text{R}}}\sin \theta \right]  \notag \\
&+& 2G_{1}+\frac{N\hbar \omega _{x}}{2}+NE_{\text{R}}.
\end{eqnarray}%
Further minimization of $E(\theta ,\alpha )$ with respect to ${\theta }$
leads to
\begin{equation}
-2\cos \theta ((1- \frac{G_{3}}{NE_{\text{R}}})\sin \theta +\frac{\Omega }{%
4E_{\text{R}}})=0.  \notag
\end{equation}%
Defining $\Omega _{\text{c}}=4E_{\text{R}}(1-G_3/NE_\text{R})$, we obtain
two different regions:

1) $\Omega >\Omega _{\text{c}}$: there is only one solution that minimizes
the mean field energy: $\cos {\theta }=0$ and $\sin {\theta }=-1$. In this
case, the spin polarization is zero, and the corresponding phase is the
spin-balanced normal phase.

2) $\Omega <\Omega _{\text{c}}$: the energy is minimized for $\sin {\theta }%
=-\Omega /[4E_{\text{R}}(1-G_{3}/NE_\text{R})]\approx -\Omega /4E_{\text{R}}$
, and there are two possible values $\cos {\theta }\approx \pm \sqrt{1-(
\Omega/4E_{\text{R}})^{2}}$, corresponding to a BEC occupying the left or
right band minimum.

\textbf{Ground state properties in the Dicke model:} In the following we
neglect the constant mean field energy terms and the small $G_{3}$ terms,
therefore $\Omega _{\text{c}}\approx 4E_{\text{R}}$. In the region $\Omega
>\Omega _{\text{c}}$, the system is in the normal phase and $v=u=0$. The
mean photon number is%
\begin{equation}
\bar{n}_{\text{photon}}=|\alpha |^{2}=u^{2}+v^{2}=0.  \notag
\end{equation}%
The spin polarizations become
\begin{eqnarray}
\frac{\langle {J_{z}}\rangle }{j\hbar } &=&-\cos \theta =0  \notag \\
\frac{\langle {J_{x}}\rangle }{j\hbar } &=&\sin \theta =-1.  \notag
\end{eqnarray}%
The ground state energy per particle is
\begin{equation*}
E_\text{g}=\frac{E}{N}=-\frac{1}{2}\Omega +E_{\text{R}}+\frac{2G_{1}}{N}+ \frac{1}{2}\hbar \omega _{x} \text{.}
\end{equation*}
We denote the first two terms as the single particle energy $E_\text{s}$.

In the region $\Omega <\Omega _{\text{c}}$, the system is in the
superradiant phase with%
\begin{equation}
\cos \theta \approx \pm \sqrt{1-(\frac{\Omega }{4E_{\text{R}}})^{2}}  \notag
\end{equation}%
and the mean photon number
\begin{equation}
\bar{n}_{\text{photon}}=v^{2}\approx \frac{\hbar k_{\text{R}}^{2}}{2m\omega
_{x}}\left( 1-\frac{\Omega ^{2}}{16E_{\text{R}}^{2}}\right) =\frac{\hbar }{%
2m\omega _{x}}q^{2} \text{,}
\end{equation}%
where $q=\pm k_{\text{R}}\sqrt{1-(\Omega/4E_\text{R})^2}$ is the
quasi-momentum of the BEC. Note that the average photon number depends on
the trapping frequency and in the main text we rescale it to $N_{\text{p}}=
2m\omega _{x} \bar{n}_{\text{photon}}/\hbar\approx q^{2}$. The spin
polarization in the superradiant phase is
\begin{eqnarray}
\frac{\langle {J_{z}}\rangle }{j\hbar } &=&-\cos \theta \approx \pm \sqrt{1-%
\frac{\Omega ^{2}}{16E_{\text{R}}^{2}}}  \notag \\
\frac{\langle {J_{x}}\rangle }{j\hbar } &=&\sin \theta \approx -\frac{\Omega
}{4E_{\text{R}}}.  \notag
\end{eqnarray}%
Finally, the ground state energy per particle is
\begin{equation*}
E_\text{g}=\frac{E}{N}\approx -\frac{\Omega ^{2}}{16E_{\text{R}}}+\frac{2G_{1}}{N}+\frac{1}{2}\hbar \omega _{x} \text{.}
\end{equation*}%
We denote the first term as the single particle energy $E_\text{s}$. The
energy expressions for $\Omega<4E_\text{R}$ and $\Omega>4E_\text{R}$ are
consistent, they are continuous at $\Omega =4E_{\text{R}}$. \newline

\textbf{Author contributions} C.H., Y.Z., C.Z. and P.E. conceived the
experiment and theoretical modeling, C.H., J.C. and P.E. performed the
experiments, C.Q., Y.Z., M.G. performed the theoretical calculations, C.Z.
and P.E. supervised the project.

\textbf{Acknowledgement} C.H., J.C. and P.E. acknowledge funding from NSF and
ARO. C.Q., Y.Z., M.G. and C.Z. are supported by ARO (W911NF-12-1-0334),
DARPA-YFA (N66001-10-1-4025), AFOSR (FA9550-13-1-0045), and NSF-PHY
(1249293). M.G. is supported by Hong Kong RGC/GRF Projects (No. 401011 and
No. 2130352) and the Chinese University of Hong Kong (CUHK) Focused
Investments Scheme.

\textbf{Competing financial interests} The authors declare no competing
financial interests.


\begin{thebibliography}{99}
\bibitem{Bloch1} Bloch, I., Dalibard, J. \& Nascimb\`{e}ne. S. Quantum
simulations with ultracold quantum gases. \textit{Nature Physics} \textbf{8}%
, 267-276 (2012).

\bibitem{Bloch2} Bloch, I., Dalibard, J. \& Zwerger. W. Many-body physics
with ultracold gases. \textit{Rev. Mod. Phys.} \textbf{80}, 885-964 (2008).

\bibitem{Ian} Lin, Y.-J., Garcia, K. J. \& Spielman, I. B.
Spin-orbit-coupled Bose-Einstein condensates. \textit{Nature} \textbf{471,}
83-86 (2011).

\bibitem{Pan} Zhang, J.-Y. \textit{et al.} Collective dipole oscillation of
a spin-orbit coupled Bose-Einstein condensate. \textit{Phys. Rev. Lett.}
\textbf{109,} 115301 (2012).

\bibitem{Peter} Qu, C., Hamner, C., Gong, M., Zhang, C. \& Engels, P.
Observation of Zitterbewegung in a spin-orbit coupled Bose-Einstein
condensate. \textit{Phys. Rev. A} \textbf{88}, 021604(R) (2013).

\bibitem{Chen} Olson, A. J. \textit{et al.} Tunable Landau-Zener transitions
in a spin-orbit coupled Bose-Einstein condensate, Preprint at
http://arXiv.org/labs/1310.1818 (2013).

\bibitem{Zhang} Wang, P. \textit{et al.} Spin-orbit coupled degenerate Fermi
gases. \textit{Phys. Rev. Lett.} \textbf{109,} 095301 (2012).

\bibitem{Martin} Cheuk, L. W. \textit{et al.} Spin-Injection spectroscopy of
a spin-orbit coupled Fermi gas. \textit{Phys. Rev. Lett.} \textbf{109,}
095302 (2012).

\bibitem{Ian4} Williams, R. A., Beeler, M. C., LeBlanc, L. J.,
Jimenez-Garcia, K. \& Spielman, I. B. A Raman-induced Feshbach resonance in
an effectively single-component Fermi gas. Phys. Rev. Lett. \textbf{111},
095301 (2013).

\bibitem{Bloch3} Aidelsburger, M. \textit{et al.} Realization of the
Hofstadter Hamiltonian with ultracold atoms in optical lattices. \textit{%
Phys. Rev. Lett.} \textbf{111}, 185301 (2013).

\bibitem{Ketterle} Miyake, H., Siviloglou, G. A., Kennedy, Burton, W. C. \&
Ketterle, W. Realizing the Harper Hamiltonian with laser-assisted tunneling
in optical lattices. \textit{Phys. Rev. Lett.} \textbf{111}, 185302 (2013).

\bibitem{Ian2} Beeler, M. C. \textit{et al.} The spin Hall effect in a
quantum gas. \textit{Nature} \textbf{498}, 201-204 (2013).

\bibitem{Ian3} Galitski, V. \& Spielman, I. B. Spin-orbit coupling in
quantum gases. \textit{Nature} \textbf{494}, 49-54 (2013).

\bibitem{Kane} Hasan, M. Z. \& Kane, C. L. Colloquium: topological
insulators. \textit{Rev. Mod. Phys.} \textbf{82}, 3045-3067 (2010).

\bibitem{Qi} Qi, X.-L. \& Zhang, S.-C. Topological insulators and
superconductors. \textit{Rev. Mod. Phys.} \textbf{83}, 1057-1110 (2011).

\bibitem{Zhai} Wang, C., Gao, C., Jian, C.-M. \& Zhai, H. Spin-orbit coupled
spinor Bose-Einstein condensates. \textit{Phys. Rev. Lett.} \textbf{105},
160403 (2010).

\bibitem{Wu} Wu, C.-J., Mondragon-Shem, I. \& Zhou, X.-F. Unconventional
Bose-Einstein condensations from spin-orbit coupling. \textit{Chin. Phys.
Lett.} \textbf{28}, 097102 (2011).

\bibitem{Ho2} Ho, T.-L. \& Zhang, S. Bose-Einstein condensates with
spin-orbit interaction. \textit{Phys. Rev. Lett.} \textbf{107}, 150403
(2011).

\bibitem{Sinha} Sinha, S., Nath, R. \& Santos, L. Trapped two-dimensional
condensates with synthetic spin-orbit coupling. \textit{Phys. Rev. Lett.}
\textbf{107}, 270401 (2011).

\bibitem{Yongping2} Zhang, Y., Mao, L. \& Zhang, C. Mean-field dynamics of
spin-orbit coupled Bose-Einstein condensates. \textit{Phys. Rev. Lett.}
\textbf{108}, 035302 (2012).

\bibitem{Hu} Hu, H., Ramachandhran, B., Pu. H. \& Liu, X.-J. Spin-orbit
coupled weakly interacting Bose-Einstein condensates in harmonic traps.
\textit{Phys. Rev. Lett.} \textbf{108}, 010402 (2012).

\bibitem{Stringari} Li, Y. Pitaevskii, L. P. \& Stringari, S. Quantum
tricriticality and phase transitions in spin-orbit coupled Bose-Einstein
condensates. \textit{Phys. Rev. Lett.} \textbf{108}, 225301 (2012).

\bibitem{Dicke} Dicke, R. H. Coherence in spontaneous radiation processes.
\textit{Phys. Rev.} \textbf{93}, 99-110 (1954).

\bibitem{Esslinger} Ritsch, H., Domokos, P., Brennecke, F. \& Esslinger, T.
Cold atoms in cavity-generated dynamical optical potentials. \textit{Rev.
Mod. Phys.} \textbf{85}, 553-601 (2013).

\bibitem{Hepp} Hepp, K. \& Lieb, E. H. On the superradiant phase transition
for molecules in a quantized radiation field: the dicke maser model. \textit{%
Ann. Phys.} \textbf{76}, 360 (1973).

\bibitem{Wang} Wang, Y. K. \& Hioe, F. T. Phase transition in the Dicke
model of superradiance. \textit{Phys. Rev. A} \textbf{7}, 831 (1973).

\bibitem{Gross} Gross, M. \& Haroche, S. Superradiance: An essay on the
theory of collective spontaneous emission. \textit{Physics Reports} \textbf{%
93}, 301-396 (1982).

\bibitem{Baumann} Baumann, K., Guerlin, C., Brennecke, F. \& Esslinger, T.
Dicke quantum phase transition with a superfluid gas in an optical cavity.
\textit{Nature} \textbf{464}, 1301-1306 (2010).

\bibitem{Ho} Ho, T.-L. Spinor Bose condensates in optical traps. \textit{%
Phys. Rev. Lett.} \textbf{81}, 742-745 (1998).

\bibitem{Bloch4} Widera, A. \textit{et al.} Precision measurement of
spin-dependent interaction strengths for spin-1 and spin-2 $^{87}$Rb atoms.
\textit{New J. Phys.} \textbf{8}, 152 (2006).

\bibitem{StamperKurn} Higbige, J. \& Stamper-Kurn D. M. Generating
macroscopic quantum-superposition states in momentum and internal-state
space from Bose-Einstein condensates with repulsive interactions. \textit{%
Phys. Rev. A} \textbf{69}, 053605 (2004).

\bibitem{Yongping} Zhang, Y., Chen, G. \& Zhang, C. Tunable spin-orbit
coupling and quantum phase transition in a trapped Bose-Einstein condensate.
\textit{Sci. Rep.} \textbf{3}, 1937 (2013).

\bibitem{Emary} Emary, C. \& Brandes, T. Quantum chaos triggered by
precursors of a quantum phase transition: the Dicke model. \textit{Phys.
Rev. Lett.} \textbf{90}, 044101 (2003).

\bibitem{Emary2} Emary, C. \& Brandes, T. Chaos and the quantum phase
transition in the Dicke model. \textit{Phys. Rev. E} \textbf{67}, 066203
(2003).

\bibitem{Bakemeier} Bakemeier, L., Alvermann, A., \& Fehske, H. Quantum
phase transition in the Dicke model with critical and noncritical
entanglement. \textit{Phys. Rev. A} \textbf{85}, 043821 (2012).

\bibitem{Chin} Parker, C. V., Ha, L.-C. \& Chin, C. Direct observation of
effective ferromagnetic domains of cold atoms in a shaken optical lattice.
\textit{Nature Physics} 9, 769 (2013).

\bibitem{Sachdev} Sachdev, S. \textit{Quantum phase transitions} (Cambridge
University Press, Cambridge, England) (1999).

\bibitem{Stringari2} Stringari, S. Collective excitations of a trapped
Bose-condensed gas. \textit{Phys. Rev. Lett.} \textbf{77}, 2360-2363 (1996).

\bibitem{Emary3} Emary, C. \& Brandes, T. Phase transitions in generalized
spin-boson (Dicke) models. \textit{Phys. Rev. A} \textbf{69}, 053804 (2004).

\bibitem{truscott} Dedman, D. J., Dall, R. G., Byron, L. J. \& Truscott, A.
G. Active cancellation of stray magnetic fields in a Bose-Einstein
condensation experiment. \textit{Rev. Sci. Instrum.} \textbf{78}, 024703
(2007).

\bibitem{Pan2} Zhang, J.-Y. \textit{et al.} Experimental determination of
the finite-temperature phase diagram of a spin-orbit coupled Bose gas.
\textit{Nature Physics} doi:10.1038/nphys2905, (2014).
\end{thebibliography}
\end{document}